# DIVERSE CORRELATION STRUCTURES IN GENE EXPRESSION DATA AND THEIR UTILITY IN IMPROVING STATISTICAL INFERENCE[1]

By Lev Klebanov and Andrei Yakovlev

*Charles University and University of Rochester*

It is well known that correlations in microarray data represent a serious nuisance deteriorating the performance of gene selection procedures. This paper is intended to demonstrate that the correlation structure of microarray data provides a rich source of useful information. We discuss distinct correlation substructures revealed in microarray gene expression data by an appropriate ordering of genes. These substructures include stochastic proportionality of expression signals in a large percentage of all gene pairs, negative correlations hidden in ordered gene triples, and a long sequence of weakly dependent random variables associated with ordered pairs of genes. The reported striking regularities are of general biological interest and they also have far-reaching implications for theory and practice of statistical methods of microarray data analysis. We illustrate the latter point with a method for testing differential expression of nonoverlapping gene pairs. While designed for testing a different null hypothesis, this method provides an order of magnitude more accurate control of type 1 error rate compared to conventional methods of individual gene expression profiling. In addition, this method is robust to the technical noise. Quantitative inference of the correlation structure has the potential to extend the analysis of microarray data far beyond currently practiced methods.

**1. Introduction.** Recent years have seen a growing interest in correlations between gene expression levels in the literature on statistical methodologies for microarray analysis [Jaeger, Sengupta and Ruzzo (2003), Szabo et al. (2002, 2003), Xiao et al. (2004), Goerman, van de Geer, de Kort and van Houwelingen (2004), Geman, d'Avignon, Naiman and Winslow (2004), Lai, Wu, Chen and Zhao (2004), Shedden and Taylor (2004), Dettling,

Received March 2007; revised May 2007.
[1]Supported by NIH/NIGMS Grant GM 075299 and J&J Discovery Concept Award.
Supplementary material available at http://imstat.org/aoas/supplements

Gabrielson and Parmigiani (2005), Lu, Liu and Deng (2005), Qiu et al. (2005a, 2005b), Almudevar et al. (2006), Klebanov, Jordan and Yakovlev (2006a), Qiu and Yakovlev (2006, 2007) Efron (2007), Gordon, Glazko, Qiu and Yakovlev (2007)]. There are at least two reasons for this tendency.

First, it has become clear that building methods of significance testing on the assumption of independence (or weak dependence) between gene expression signals is in conflict with the fact that correlations in microarray data are not only very strong but they are also long-ranged, involving thousands and even tens of thousands of genes. The consequences of ignoring this fact may be disastrous, manifesting themselves in highly unstable results of estimation or testing [Qiu, Kleband and Yakovlev (2005b), Klebanov and Yakovlev (2006), Qiu and Yakovlev (2006)]. All multiple testing procedures are very unstable in the presence of correlations between gene expression levels which is the main factor causing instability of gene lists [Qiu, Xiao, Gordon and Yakovlev (2006)]. As a result, the actual number of false discoveries is not well controlled even if strong control is guaranteed in terms of expected values. The variances of type 1 and type 2 errors appear to be no less important, as performance indicators of a given multiple testing procedure, than their respective mean values.

Second, it has gradually come into the realization that inter-gene correlations are not just a nuisance, complicating statistical inference from microarray gene expression data, but a rich source of useful information. As larger sets of microarray data become more readily available, methodological studies focused on stochastic dependencies between gene expression levels are gaining in importance.

In the present paper we describe some striking regularities manifesting themselves in real microarray data. These regularities have to do with distinct correlation structures which can be revealed by an appropriate ordering of observed gene expressions. While such structures still lack a clear biological interpretation, they represent stable mass phenomena and provide an additional source of information for various purposes of data analysis. The inference of stochastic dependencies in gene sets has the potential to significantly extend and enrich the approaches focused solely on marginal distributions of gene expression signals and associated two-sample tests. However, this research avenue is still in its infancy and much more empirical knowledge needs to be amassed before a solid methodological foundation can be built for conceptually novel approaches to microarray data analysis. Our findings in the present paper enable us to progress in this direction.

Yet another way to benefit from studies of correlations in microarray data is to find a structure that yields a long sequence of weakly dependent random variables. One such structure is described in the present paper and its usefulness is demonstrated with a method for selecting differentially expressed genes in nonoverlapping gene pairs. This method is an order of magnitude more accurate in controlling type 1 error rate compared to conventional methods of individual gene expression profiling. The proposed method also offers a more sensitive indicator of changes in gene expression patterns than two-sample test-statistics based on expression levels.

In accordance with the above comments, the objective of the present paper is two-fold:

1. To introduce the biostatistician into the captivating realm of diverse correlation substructures in microarray gene expression data, some of which may have important biological implications.

2. To discuss basic facts about a sequence of weakly dependent random variables deciphered in several data sets and a possible way of exploiting its properties to improve statistical inference from high-density oligonucleotide microarray data.

The paper is organized as follows. In Sections 2 and 3 we describe special types of correlation revealed in pairs and triples of genes. A special emphasis is placed on a valid concern about the contribution of technical noise and cross-hybridization to observed phenomena. These examples show that the actual correlation structure of gene expression data is by far more complex than that traditionally modeled in simulation studies designed to validate various inferential procedures. The information on "differential dependence" in different phenotypes can be utilized to make the selection of candidate genes more meaningful (Section 2). We also believe that the deciphered types of stochastic dependence hold significance in understanding the modes of gene regulation at the transcription level. The reader who is more interested in statistical aspects of the problem can skip these two sections without any loss of information.

In Sections 4 and 5 we introduce a newly uncovered sequence, referred to as the $\delta$-sequence, of weakly dependent random variables and study its properties. All methods used in this study are nonparametric and do not involve any modeling. The $\delta$-sequence was first found and characterized in the HYPERDIP data (Table 1). To confirm its universal character, we later interrogated several other data sets produced by the Affymetrix technology and listed in Table 1. Since our analysis was to a large extent based on correlation coefficients, the choice of data sets in Table 1 was governed by the quest for larger sample sizes. Sequences with properties similar to those of the $\delta$-sequence in the HYPERDIP data were proven to be present in all data sets we have studied so far.

TABLE 1
*Data sources. PCMIT: paired samples of tumor and normal tissues, BC: samples obtained from untreated breast cancer patients, BCCL: breast cancer cells cultured in vitro (only "vehicle" control samples that were treated with the medium used to solubilize the inhibitor), TELL and HYPERDIP: two types of childhood leukemia, CECL: mouse colon epithelium cells cultured in vitro. The Affymetrix platform was used in all studies*

| # | Tissue source | Code | Sample size | Reference |
|---|---|---|---|---|
| 1 | Prostate cancer | PCMIT | 52 | Singh et al. (2002) |
| 2 | Breast cancer | BC | 61 | Sotiriou et al. (2006) |
| 3 | Breast cancer cell line | BCCL | 47 | Lamb et al. (2006) |
| 4 | TELL childhood leukemia | TELL | 79 | Yeoh et al. (2002) |
| 5 | HYPERDIP childhood leukemia | HYPERDIP | 88 | Yeoh et al. (2002) |
| 6 | Colon epithelium cell line | CECL | 20 | Klebanov et al. (2006b) |

The $\delta$-sequence appears to be a very stable structure as far as the property of weak dependence between its elements is concerned. This finding immediately suggests a new paradigm of microarray data analysis, namely, a search for differentially expressed genes in nonoverlapping gene pairs. While not equivalent to the commonly accepted concept of individual gene profiling, the new paradigm has certain advantages over the traditional approach, as discussed in Section 6. However, many methodological avenues have yet to be explored before this paradigm gives rise to a practically useful and widely accepted method of microarray data analysis.

**2. Stochastic proportionality in gene pairs.** It is a well-known fact that pairwise correlations between gene expression levels are overwhelmingly positive and strong [Qiu et al. (2005a, 2005b), Almudevar et al. (2006)]. When studying various data sets, we observed the average (over all gene pairs) of correlation coefficients ranging from 0.84 to 0.97. Another feature of the correlation structure of microarray data is that the correlations are long-ranged, that is, a particular gene may have very high correlation coefficients with a vast majority of other genes. A typical example of this situation is shown in Figure 1 presenting a histogram of correlation coefficients in all gene pairs formed by a particular gene in PCMIT. The mean (over the genes) value of the histogram is 0.78, while the corresponding standard deviation is equal to 0.16. Such long-range strong correlations prevail in a huge proportion of randomly selected genes. To make the analysis of correlations tractable, it is important to identify stable correlation patterns (substructures) that are either universal or specific to the phenotype under study.

In a recent paper [Klebanov, Jordan and Yakovlev (2006a)], we described a special type of stochastic dependence between expression levels in pairs of genes. This modulation-like *unidirectional* dependence between expression signals was discerned in three large sets of microarray data on childhood leukemia and later confirmed by a similar analysis of some other data sets. A distinctive feature of this type of dependence, termed the type A dependence, is that the expression of a "gene-modulator" is stochastically proportional to that of a "gene-driver." To provide a formal definition of type A dependence, let $g_x$ and $g_y$ be two genes in a given pair, and let random variables $X$ and $Y$ represent their respective expression levels. A pair of genes $(g_x, g_y)$ is said to be type A if $X$ and $Y$ satisfy the condition

$$Y = XZ, \tag{1}$$

where $Z$ is a positive random variable stochastically independent of $X$. In this case, we say that $g_x$ is a driver, while $g_y$ is a modulator. The reason for introducing this terminology is that the roles of $g_x$ and $g_y$ in a type A pair are not symmetrical because, given (1) is true, the random variables $Y$ and $1/Z$ are no longer independent; this is precisely what makes the type A stochastic dependence so special. All other stochastic dependencies in gene pairs are operationally classified as type B dependence.

By log-transforming expression (1), we have

$$y = x + z, \tag{2}$$

where $x = \log X, y = \log Y, z = \log Z$. A general necessary condition for the type A dependence is

$$\mathrm{Var}(x) = \mathrm{Cov}(x, y).$$

This condition is sufficient under joint normality of $(x, y)$. It also follows from (2) that the type A dependence induces an ordering of the random variables $x$ and $y$ in terms of their variances because $\mathrm{Var}(x) < \mathrm{Var}(y)$ in each type A pair.

The main concern in studies of the correlation structure of microarray data is the presence of a multiplicative technical noise that can induce spurious correlations. Let $x_{ij}, i = 1, \ldots, m, j = 1, \ldots, n$, be the true log-expression of the $i$th gene on the $j$th array, where $m$ is the total number of genes and $n$ is the number of arrays. The unobservable random signals $x_{ij}$ are expected to be highly variable due to the inherent biological (inter-subject) variability. A general log-linear noise model is

$$x_{ij}^* = x_{ij} + \epsilon_{ij}, \tag{3}$$

where $x_{ij}^*$ is the observable log-expression level and $\epsilon_{ij}$ is a log-additive independent random measurement error (noise). This model is clearly unidentifiable as no inference of $x_{ij}$ (or the noise component $\epsilon_{ij}$) is possible from the observed signal $x_{ij}^*$. A particular case of (1) given by

$$x_{ij}^* = x_{ij} + \epsilon_j \tag{4}$$

is the most widely accepted array-specific random effect model. However, the reduced model (4) is generally unidentifiable as well. While the structure of (4) looks similar to that of (2), the main difference between them is that $z$ affects only one member in a type A gene pair and varies in magnitude among such pairs, while $\epsilon_j$ is the same for all genes on a given array.

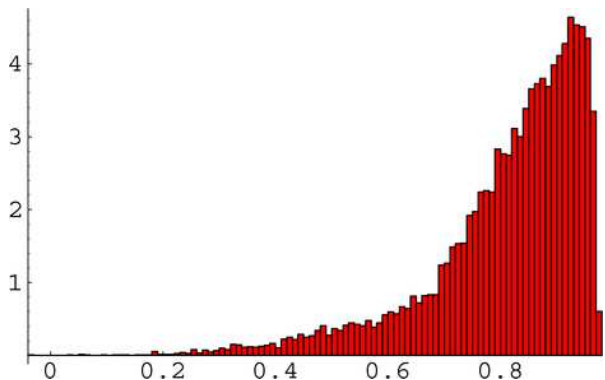

FIG. 1. *Histogram of correlation coefficients computed for all gene pairs formed by gene IPF1 in PCMIT. The selected gene encodes a transcription factor involved in regulation of transcription and morphogenesis.*

REMARK 1. The existence of type A dependence provides an alternate proof of the fact that the true expression signals are correlated, regardless of whether or not the log-additive technical noise is present. Consider two noisy log-expressions $x_1^*$ and $x_2^*$ and assume that the true signals $x_1$ and $x_2$ represent nondegenerate random variables. If the pair $x_1^*, x_2^*$ demonstrates type A dependence, it follows that $x_1 + \epsilon$ and $x_2 - x_1$ are independent, that is, $\text{Cov}(x_1 + \epsilon, x_2 - x_1) = 0$. Since $\epsilon$ is assumed to be independent of $x_1$ and $x_2$, $\text{Cov}(x_1, x_2 - x_1) = 0$ as well. The latter condition implies that $x_1$ and $x_2$ are correlated. Indeed, suppose the reverse is true, that is, $x_1$ and $x_2$ are independent. Then $\text{Cov}(x_1, x_2 - x_1) = -\text{Var}(x_1)$ which can only be true if $x_1$ is a degenerate random variable, a contradiction.

A recently published report of the MicroArray Quality Control (MAQC) Consortium [Shi et al. (2006)] provides direct measurements of the technical noise specific for the Affymetrix platform in the absence of biological variability. Our re-analysis of the MAQC study [Klebanov and Yakovlev (2007)] revealed two important features of the technical noise. First, the standard deviations of the random component of technical noise appear to be symmetrically distributed across genes and this distribution is quite narrow. In other words, the reduced model given by formula (4) is in good agreement with the MAQC data generated from technical replicates. Second, the observed level of technical noise is quite low and its effect on correlation coefficients estimated from microarray data is negligibly small [Klebanov and Yakovlev (2007)].

Klebanov, Jordan and Yakovlev (2006a) estimated the abundance of type A pairs among all gene pairs formed from a total of 12558 genes (probe sets).

They used the HYPERDIP data and proceeded from the noise model given by formula (4). A statistical test was intentionally designed to select only those pairs that demonstrate the type A dependence beyond the random effect attributable to technical noise. A total of more than 35% of all pairs formed from 12558 genes were conservatively estimated to belong to type A in the data sets under study. It later transpired that the above estimate was overly conservative; utilizing estimates of the noise level resulted from our analysis of MAQC data indicates that the proportion of type A pairs for HYPERDIP is closer to 50%.

Yet another possible concern is that probes with a similar sequence might display correlation simply because they are cross-hybridizing to the same transcript, thereby inducing many groups of genes that look like type A. Klebanov and Yakovlev (2007) argued that the effects of cross-hybridization are probably not very disturbing just because they do not manifest themselves in a high level of technical noise. Multiple targeting, however, is a complex and essentially unobservable process so that its distorting effect cannot be ruled out even if the overall technical noise is low. In Supplementary Material 1 we consider a simple stochastic model of cross-hybridization to demonstrate that the correlation between two independent probe-sets targeting the same transcript does not need to be positive. We also discuss a recent paper by Okoniewski and Miller (2006) concerned with adverse effects of multiple targeting on the correlation structure of Affymetrix data. A more comprehensive discussion of this issue will be given in a forthcoming paper. Our conclusion is that there is no compelling reason to attribute the phenomenon of type A dependence to the effects of cross-hybridization.

As discussed in the paper by Klebanov, Jordan and Yakovlev (2006a), the modulation-like dependence, manifesting itself in type A gene pairs, is a stable mass phenomenon. When studying this phenomenon, the authors noticed that type A pairs form long branching chains such that each driver is at the same time a modulator of another (upstream) driver. Such chains can involve thousands of genes. The composition of a particular chain may undergo dramatic changes across two phenotypes. In an effort to endow the uncovered types of stochastic dependence with a biological meaning, Klebanov, Jordan and Yakovlev (2006a) looked closely at annotations of those genes that were declared differentially expressed by nonparametric significance testing in a comparison of two types of leukemia. The authors noted that modulators tend to be among genes with the largest differences in expression levels. The subset of modulators was preferentially enriched with signal transduction factors, while the set of genes forming type B pairs in both phenotypes under comparison was enriched with transcription factors.

Additional sources of information are necessary to make selection of genes based on observed $p$-values more meaningful and less biased toward the most pronounced differential expression. It should always be kept in mind that the magnitude of differential expression does not imply biological significance, while modern methods of microarray data analysis are biased toward the strongest effects due to the nature of multiple testing. The above example demonstrates that quantitative insights into the correlation structure have the potential to provide such sources of information and extend the analysis of microarray data far beyond the currently practiced methods focused solely on differentially expressed individual genes.

**3. Negative correlations in ordered gene triples.** Since the publication by Klebanov, Jordan and Yakovlev (2006a), another interesting phenomenon has been discovered in gene triples. As was mentioned in the previous section, the between-gene correlations are overwhelmingly positive. A substantial part of strong positive correlations is contributed by type A dependence in numerous gene pairs. Consider now a *triple* of genes formed by two type A pairs. In such a triple, the vector of expressions with their variances arranged in increasing order is of the form $(x, x + z_1, x + z_1 + z_2)$, where $z_1$ is independent of $x$ and $z_2$ is independent of $x + z_1$. Therefore, we have the relationship

$$\mathrm{Cov}(x, z_2) = - \mathrm{Cov}(z_1, z_2),$$

that is, the covariance between the increments $z_1$ and $z_2$ is expected to be negative whenever $\mathrm{Cov}(x, z_2) > 0$. How frequently the latter condition is met can be assessed only with real data. To answer this question, we resorted to the HYPERDIP data set.

From randomly chosen genes, we formed 30,000 triples of genes arranged in increasing order of their variances and composed exclusively of type A pairs. Much to our surprise, more than 95% of such triples of genes were characterized by *negative* values of $\mathrm{Cov}(z_1, z_2)$ in the biological data under study. Next we randomly formed 500,000 triples of genes ordered in the same way but composed of *any* gene pairs. In this case, over 88% of triples had negative values of $\mathrm{Cov}(z_1, z_2)$.

REMARK 2. It should be noted that the inference of correlations between $z_1$ and $z_2$ is invariant to a multiplicative (log-additive) random measurement error. Indeed, the most widely accepted random effect model (4) assumes that the same log-additive measurement error is shared by all genes on each array, but the level of this error varies randomly from array to array. In the presence of this random effect denoted by $\epsilon$, we observe $(x + \epsilon, y + \epsilon)$, where $\epsilon$ is a positive random variable independent of the pair $(x, y)$. For any joint realization of $x, y, z_1, z_2, \epsilon$, it is clear that $z_1 = (y + \epsilon) - (x + \epsilon) = y - x$, so that $z_1$ is a noise-free random variable. Similarly, one can write $z_2 = (y + z_2 + \epsilon) - (y + \epsilon) = y + z_2 - y$ to establish the same fact for $z_2$.

To better appreciate the meaning of the negative correlation observed in ordered triples of genes, one can resort to the following loose argument. Assuming an approximately linear relationship between $z_1$ and $z_2$, notice that the rate of change (increment) in values of $y + z_2$ relative to $y$ is roughly characterized by the slope in the regression of $y + z_2$ on $y$. For every fixed value of $x$, this slope is bounded from above by 1 whenever $\mathrm{Cov}(z_1, z_2)$ is negative. This line of reasoning makes the relevance of the uncovered types of stochastic dependence to the regulation of gene function more plausible. It is not unreasonable to hypothesize that the negative correlation observed in ordered gene triples is a manifestation of mechanisms designed to stabilize chains of type A gene pairs. Another plausible hypothesis is that regulatory mechanisms of gene expression are sensitive to rates (increments) rather than to current concentrations of gene products in the cell. However, more specific biological implications of the observed phenomenon have yet to be explored in experiments that provide the necessary information on causal relationships in triples of genes ordered by the variance of their expressions. When looking solely at snapshots of gene expression produced by microarrays, one cannot make inference on the interactions between genes far beyond the scope of correlation analysis. To better understand the regulatory role of the increments $z_1$ and $z_2$, more information needs to be extracted from gene perturbation experiments, which may be an interesting research avenue in gene regulatory network reconstruction.

There might be a natural concern that the observed negative correlation between the increments $z_1$ and $z_2$ in those triples of genes that are composed of type A pairs is driven by the specific structure of such triples. However, the model behind the type A dependence is flexible enough to allow for any sign (negative or positive) of the covariance $\mathrm{Cov}(z_1, z_2)$ between the increments $z_1$ and $z_2$. This point is illustrated by the following example. Consider three random variables $u, a, b$ having a joint normal distribution. Then the covariance matrix of the form

$$(5) \quad \begin{bmatrix} \sigma_u^2 & 0 & -\mathrm{Cov}(a,b) \\ 0 & \sigma_a^2 & \mathrm{Cov}(a,b) \\ -\mathrm{Cov}(a,b) & \mathrm{Cov}(a,b) & \sigma_b^2 \end{bmatrix},$$

where $\sigma_u^2, \sigma_a^2, \sigma_b^2$ are all distinct from zero, gives a necessary and sufficient condition for the pairs of random variables $(u, v = u+a)$ and $(v, w = v+b)$ to be type A dependent. It is clear that the variances of $u, v$ and $w$ are ordered in the triple $(u, v, w)$ so that $\sigma_u < \sigma_v < \sigma_w$. Note that $\mathrm{Cov}(v-u, w-v) = \mathrm{Cov}(a, b)$ of any sign is consistent with the structure of matrix (5). In other words, for any $\mathrm{Cov}(v-u, w-v)$, there always exists a multivariate normal model for $(u, a, b)$ which is consistent with this covariance as long as the ordered triple $(u, v, w)$ is formed from two pairs of type A. Therefore, the observed abundance (more than 95% of ordered triples) of negative correlations between the above-mentioned increments is likely to be a biological phenomenon rather than a mathematical feature of the model. This may also be considered an indirect indication of deterministic relationships between the increments, but such relationships cannot be discerned in microarray data alone; they should be further studied in gene perturbation or time-course experiments.

Consider now an arbitrary (not necessarily composed of two type A pairs) triple $(u, v, w)$ ordered by the variances of its components. Let $\rho(u,v)$ and $\rho(v,w)$ be the correlation coefficients for $(u,v)$ and $(v,w)$, respectively. It is not difficult to derive a sufficient condition for $\rho(v-u, w-v)$ to be positive. This condition is of the form

$$\rho(v,w) > 1 - \frac{1}{2}\left(1 - \frac{\sigma_v}{\sigma_w}\right)^2 \left(1 - \frac{\sigma_v}{\sigma_u}\right)^2,$$

regardless of the value of $\rho(u,v)$. This condition does not seem to be too restrictive. Recall that less than 12% of all ordered triples tend to display a positive correlation between the increments.

**4. A structure yielding near-independent random variables.** In the decade since the advent of microarray technology, numerous methodological papers have been pursuing the idea of pooling either expression measures or associated test-statistics (or $p$-values) across genes in order to develop better procedures for selecting differentially expressed genes [see Qiu, Klebanov and Yakovlev (2005b), Klebanov and Yakovlev (2006) for some relevant references]. This idea still dominates the literature on microarray data analysis. Although never stated explicitly, the actual goal of such endeavors is to overcome the notorious sample size and cost limitations. The authors of all such papers pin their hopes on the fact that the number of genes $m$ is very large so that the effective sample size can be increased and the asymptotics in the sample size $n$ can be replaced with the asymptotics in the number of genes $m$. If the levels of gene expression or certain statistics built on them were independent, this expedient would be expected to work well even in small sample studies. However, this dreamy situation has little to do with the harsh reality of microarray data; their actual correlation structure (see Section 2) stands in the way of those methods that resort to pooling across genes. If one chooses to ignore this fact and applies a method proven to perform well under the assumption of independence, the price to be paid will be an intolerably high variance of the number of type 1 errors as well as the total number of rejected hypotheses [Qiu, Klebanov and Yakovlev (2005b)]. The same is true for estimation of the false discovery rate (FDR) from heavily dependent $p$-values [Qiu and Yakovlev (2006)]. The effective sample size appears to be much smaller, and not larger, when considering the pooled expression measurements as a sample [Klebanov and Yakovlev (2006)] and consistency of various estimators as functions of $m$ does not hold true [Qiu et al. (2005a), Klebanov and Yakovlev (2006)]. The relevance of correlations to the validity of a much used formula for the FDR estimation is demonstrated in Qiu and Yakovlev (2007).

As far as significance testing for differential expression is concerned, it is obviously the violation of independence assumptions, rather than dissimilar distributions of expression levels for different genes, that represents the main hurdle, especially when using distribution-free test-statistics. There have been attempts to generalize methods based on independence by thinking of gene expression measurements as a sequence of weakly dependent random variables [Storey, Taylor and Siegmund (2004)]. There are two problems with this approach. First, there is no external basis, like time or space, for structuring such a sequence because microarray technology by itself does not offer any natural way to order the genes. Second, the fact that correlations between genes are extremely strong and long-ranged leaves little hope that the assumption of any kind of weak dependence may make a difference in analysis of real biological data.

However, if we were able to find a structure in suitably transformed expression signals yielding a long sequence of near-independent random variables, this would put us in the desirable situation where methods that resort to pooling test-statistics across genes, such as adaptive FDR-controlling procedures [Benjamini and Hochberg (2000), Reiner, Yekutieli and Benjamini (2003), Storey, Taylor and Siegmund (2004)], nonparametric empirical Bayes method [Efron (2003, 2004, 2007), Efron et al. (2001)], or some robust statistical methods [Sen (2006)], may be very advantageous. This structure does not need to be externally determined; it can just as readily be data-driven and its specific make-up (i.e., gene ordering) may vary across experimental settings. However, it is desirable to have a systematic procedure for finding such structures in real data. The results produced by this procedure have to be verifiable with the data at hand and satisfy certain stability requirements. In a search for such structures, it was natural to begin by exploring the ordering of genes by values of the variance of their expression levels, just as we did in Section 3.

In what follows, use will be made of the HYPERDIP data set referred to in Sections 1 and 3. Since the original probe set definitions in these data are known to be inaccurate [Dai et al. (2005)], we updated them by using a custom CDF file from http://brainarray.mbni.med.umich.edu, which was added to the Bioconductor package to produce values of gene expressions. This procedure reduced the total number of probe sets (genes) from 12558 to 7084. The latter set is believed to be much more reliable in terms of gene identities. All our findings described in Sections 2 and 3 were confirmed with this reduced set of genes. Furthermore, the between-gene correlations appear to be, on average, stronger than in the original data set. All expression signals were log-transformed. The data were not normalized. Normalization procedures tend to induce spurious negative correlations by interfering in the joint distribution of true biological signals and, consequently, in their correlation structure. This effect was documented by Qiu et al. (2005a).

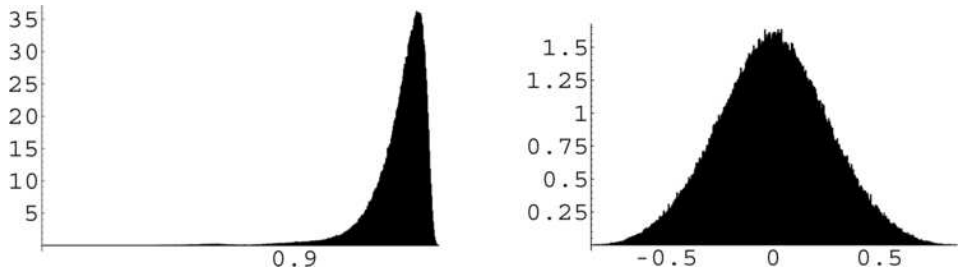

FIG. 2. *Histogram of correlation coefficients for all gene pairs formed from the first 2000 ordered genes (left panel) and for all pairs of the corresponding elements $\delta_i$ (right panel). The mean values are 0.97 for expression levels (left panel) and $-1.5 \times 10^{-4}$ for $\delta_i$ (right panel).* HYPERDIP *data set.*

When the genes are ordered by arranging their variances in increasing order, each gene is assigned a number from $i = 1$ through $i = m$, where $i = 1$ corresponds to a gene with the minimal variance, while $i = m$ corresponds to a gene with the maximal variance. Notice that $m$ is an even number in our case. If the total number of genes is odd, the last (or the first) gene in the original sequence should be discarded. We know that pairwise correlations in the sequence thus ordered tend to be very high. For example, if we select only the genes with even numbers in the ordering and estimate all pairwise correlation coefficients between their expression levels, strong positive correlations will prevail. The mean (over the selected gene pairs) value of these correlation coefficients is 0.942 with the corresponding standard deviation being equal to 0.036.

The situation changes dramatically if we form a sequence of the increments $\delta_i = x_{2i} - x_{2i-1}$ between log-expression levels of the $2i$th and the $(2i-1)$th gene $(i = 1, \ldots, m)$ in the above-defined ordering of genes. This specific sequence $\{\delta_i\}$, henceforth termed the $\delta$-sequence, generates near-independent random variables. It should be noted that the $\delta$-sequence is formed on each chip separately, while the ordering is determined by the set of chips. Figure 2 compares sample correlation coefficients between log-expression levels with those observed in pairs of the increments $\delta_i$. While in the former case the histogram of correlation coefficients is shifted to the right, it becomes symmetric around zero in the latter. Similar properties of the $\delta$-sequence versus gene expression have been confirmed in all data sets listed in Table 1 (see Supplementary Material 2).

Figure 3 displays the histogram of Fisher's transformation $z$-scores computed for all pairs of the increments $\delta_i$, $i = 1, \ldots, m/2$, in the updated HYPERDIP data. If these increments were independent, the distribution of the corresponding $z$-scores would be approximately normal with mean zero and standard deviation equaling $1/\sqrt{n-3} = 0.11$ ($n = 88$). The estimated mean and standard deviation are equal to $8 \times 10^{-5}$ and 0.35, respectively. A normal distribution with these parameter values provides a very good fit to the observed $z$-scores (Figure 3). The discrepancy between the expected and observed standard deviations suggests that elements of the sequence $\{\delta_i\}$ are not entirely independent. Insufficiency of the sample size ($n = 88$) for the normal approximation to hold may contribute to this discrepancy as well.

**5. Further properties of the sequence $\{\delta_i\}$.** Our preliminary insight into the sequence $\{\delta_i\}$ of increments between log-expression levels of the $2i$th and the $(2i-1)$th gene ($i = 1, \ldots, m$) suggests that the elements of this sequence are weakly dependent (or almost independent) random variables (Section 4). Let us now explore some other properties of $\{\delta_i\}$ that are expected to hold if our conjecture is true. These properties are also important in the assessment of practical implications of our finding, such as its utility in designing new statistical methods for microarray data analysis.

The idea of pooling expression levels or test-statistics across genes originates from the belief that the law of large numbers is met when one considers their values associated with different genes as a sample drawn from some population distribution. The observed pairwise correlations between elements of the sequence $\{\delta_i\}$ are deemed to be sufficiently low to expect statistical estimators constructed from them to be consistent (as $m \to \infty$) and have a small variance. This, in turn, suggests that pooling across observed values of the increments $\delta_i$ (rather than across the original expression levels!) may be warranted, thereby offering considerable scope for such methods as the nonparametric empirical Bayes method in finding differentially expressed gene pairs.

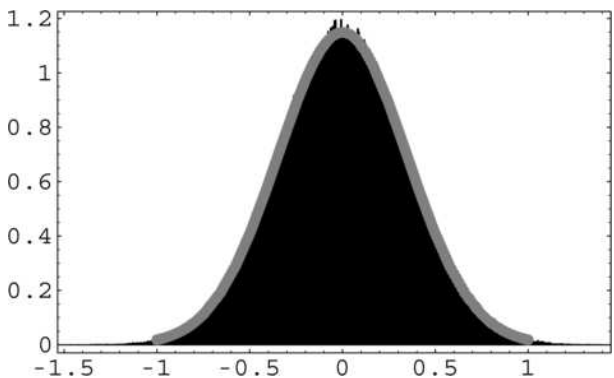

FIG. 3. *Histogram of Fisher's transformation scores computed for all pairs of increments in non-overlapping adjoining gene pairs formed from the first 2000 genes that are ordered by the variance of their expression levels. Solid line is the normal density with mean 0 and standard deviation equaling its sample counterpart computed for the population of gene pairs. HYPERDIP data set.*

A sufficient condition for the law of large numbers (Markov theorem) in this case is that the variance of the moving sample mean taken over elements of the sequence of variables under consideration tends to zero as the number of elements increases. This, in turn, guarantees consistency of the sample mean thus defined, a desirable property of any statistical estimator. To verify this condition, we computed the arithmetic mean of the first $k \times 100$ elements $\delta_i$ on each array available from the TELL data while varying the value of $k$ from 1 to 35. Next, the sample standard deviation of these mean values was computed across all arrays available from the data. The trajectory of the standard deviation shown in Figure 4 is consistent with our expectations: it tends to zero as $k$ increases. This is in drastic contrast to what is observed when the original log-expression levels associated with individual genes are pooled together in an effort to take advantage of large numbers of genes. In the latter case, the variance of the arithmetic mean of log-expression levels does not change (let alone the tendency to vanish) no matter how many genes are involved in the averaging across genes (Figure 4), a pattern consistent with our previous observations in other applications [Qiu et al. (2005a), Klebanov and Yakovlev (2006)]. The same effect was invariably observed in other data sets, Figure 5 representing another example obtained from the BC data. This shows that what is doomed to failure when pooling log-expression levels of each individual gene becomes achievable when pooling elements of the $\delta$-sequence. It should be kept in mind, however, that the increments $\delta_i$ are characteristics of gene pairs rather than individual genes.

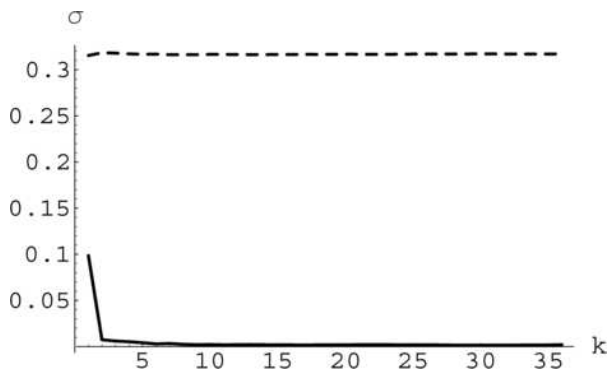

FIG. 4. *Assessing consistency of the arithmetic mean in the TELL data set ($n = 79$). Solid line: standard deviation (across arrays) of the moving mean (across a subsequence $\{\delta_i\}$, $1 \leq i \leq 100 \times k$) as a function of $k$ for elements of the sequence $\{\delta_i\}$, $i = 1, \ldots, m/2$. Dashed line: the corresponding standard deviation for the moving mean of log-expression levels for genes with even numbers in the ordering.*

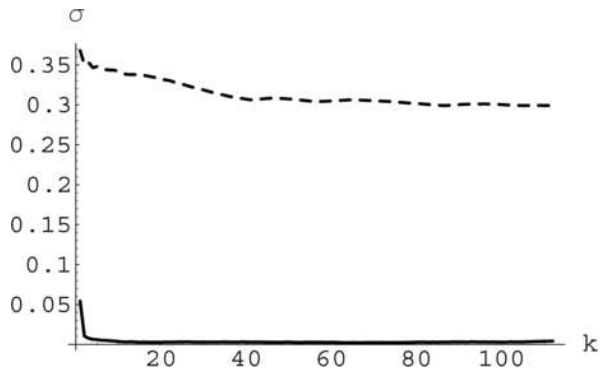

FIG. 5. *Assessing consistency of the arithmetic mean in the* BC *data set* ($n = 61$). *The same notation as in Figure 4.*

Yet another test was carried out as follows. Two subsamples were drawn without replacement from the HYPERDIP data, each containing 40 arrays. Since no difference for any gene is expected to exist between the general populations from which the two subsamples are drawn, this is a natural way to model the complete null hypothesis. The ordering of genes with respect to the variance of their log-expressions in the original sample of 88 arrays was fixed and retained the same in both subsamples irrespective of the actual values of empirical variances for different genes in each subsample. The Kolmogorov–Smirnov statistic was computed for each $\delta_i$, resulting in about 3500 values of this statistic believed to be independent and identically distributed. Recall that the Kolmogorov–Smirnov statistic is distribution-free under the null hypothesis. The prediction is that the empirical distribution function constructed from the 3500 values of the Kolmogorov–Smirnov statistic will be very close to the exact Kolmogorov–Smirnov distribution for $n_1 = n_2 = 40$. As seen from Figure 6(a), the two distributions are virtually indistinguishable. However, they become quite dissimilar when the same procedure is applied to the original log-expression levels [Figure 6(b)]. We repeated the resampling procedure many times to see that it produces remarkably stable results.

REMARK 3. Out of curiosity, we also looked at how the mean values of $\delta_i$ are distributed over the $\delta$-sequence in the HYPERDIP data. These values were obtained by averaging each $\delta_i$ over arrays (subjects). Shown in Supplementary Material 3 is a histogram of such mean values suggesting that the underlying distribution may have heavy tails.

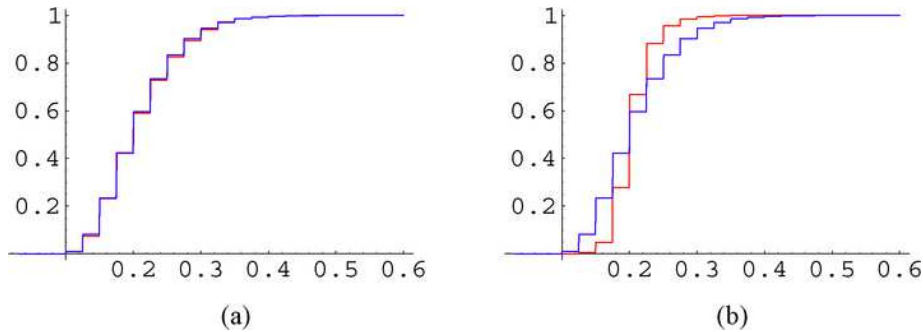

Fig. 6. *Comparison of the exact cumulative distribution function for the two-sample Kolmogorov–Smirnov statistic with the empirical distributions resulted from pooling the test-statistics associated with each $\delta_i$ across elements of the sequence $\{\delta_i\}$ (left panel) and individual gene expressions (right panel), respectively. For the $\delta$-sequence, the discrepancy between the two distributions is almost intangible, but it is quite substantial for expression levels. HYPERDIP data set.*

There is a natural concern as to how stable the proposed ordering of genes is in practical applications. Ideally, we would like to order the genes by values of their true (population) variances. Since we do not know their values, we resort to sample variances for this purpose, which may cause instability in gene ranking, especially with small/moderate sample sizes. The most fundamental question here is whether or not a certain degree of instability in gene ranking translates into instability of the correlation structure of the sequence $\{\delta_i\}$ which may ultimately be needed for practical purposes. To answer this question we conducted the following resampling study.

Our resampling procedure was based on the delete-$d$-jackknife method. In accordance with recommendations in the literature [Efron and Tibshirani (1993)], we left out $d = 11$ arrays to draw each subsample of size 77 from the HYPERDIP data consisting of 88 arrays. In each subsample, a new ordering of genes by values of the sample variance of their expression levels was determined and the corresponding sequence $\{\delta_i\}$ was constructed. Since the resampling procedure is extremely time-consuming, this study was limited to the first 150 members in the sequence $\{\delta_i\}$ identified in each subsample. In every subsample, all pairs were formed from the first 150 elements of $\{\delta_i\}$ and the corresponding pairwise Fisher's $z$-scores were computed. The $z$-scores were summarized in the form of their empirical distribution function. The procedure was repeated $B$ times, resulting in $B$ empirical distribution functions $\hat{F}_s(x)$, $1 \leq s \leq B$. Each of these functions was then compared with their arithmetic mean $\hat{G}_B(x)$ by computing the Kolmogorov distance between them. This distance should not be confused with the Kolmogorov–Smirnov statistic, as it does not include a normalizing factor and varies from 0 to 1. More specifically, the distance for a given subsample $s$, $1 \leq s \leq B$, is

given by

$$\hat{\rho}_s = \sup_z |\hat{F}_s(z) - \hat{G}_B(z)|,$$

where $z$ ranges from a minimum of observed $z$-scores to their maximum. When carried out with $B = 100$, our simulations resulted in the mean (across subsamples) value of $\hat{\rho}_s$ equaling 0.0167, while the standard deviation of the $\hat{\rho}_s$-values was equal to 0.0084. This indicates that the weak correlation structure of the sequence $\{\delta_i\}$ is well guarded against sample variations in gene ordering.

**6. Testing differential expression of nonoverlapping gene pairs.** Presented in this section is a study of the performance of a gene selection procedure based on the sequence $\{\delta_i\}$. This study is focused solely on type 1 error rates. A comprehensive assessment of the power of various multiple testing methods in conjunction with this type of microarray data analysis will be addressed in another paper.

To preserve the actual correlation structure of gene expression levels as much as possible, we carried out our study by resampling from the HYPER-DIP data. The following resampling procedure was employed to artificially produce the effects of differential expression. The original sample of size $n = 88$ was randomly divided in two parts, each containing 44 arrays. The first split was used to order the genes in one half of the original sample and estimate the standard deviations of all increments $\delta_i$, $i = 1, \ldots, 3542$. We designated 350 (randomly selected) elements of the sequence $\{\delta_i\}$ for further modification as described below. One half of the sample, henceforth referred to as subsample 1, was left intact, while a constant (across arrays) effect size was added to the pre-defined increments $\delta_i$, $i = 1, \ldots, 350$, in all arrays pertaining to the second subsample. To make the effect size reasonably invariant across different scales of the data, we used two standard deviations of $\delta_i$ (estimated from subsample 1) to specify a $\delta_i$-wise constant to be added to each increment. The distribution of the resultant effect sizes over the 350 elements of the sequence $\{\delta_i\}$ is shown in Supplementary Material 4. These values were fixed throughout the subsequent resampling cycles and the first split was discarded in the final analysis. The ordering of genes determined in subsample 1 was enforced in subsample 2.

The two subsamples thus created were used to detect "truly differentially expressed" gene pairs by the Kolmogorov–Smirnov exact test combined with an extended Bonferroni multiple testing procedure [Gordon et al. (2007)]. This procedure is known to control the per family error rate (PFER), defined as the expected number of false discoveries. By controlling the PFER, rather than the family-wise error rate (i.e., the probability of at least one type 1 error), one can gain much more power in detecting alternative hypotheses. In our study, we provided control of the PFER at a level of 9. The pre-defined PFER level is arbitrary because its specific value is immaterial for the purposes of our study. The resampling procedure was repeated 3000 times, the number of false discoveries and the apparent FDR were averaged over the subsamples, and their corresponding standard deviations were estimated. Providing a large number of subsamples is important in the presence of correlations because of huge fluctuations in the number of rejected hypotheses across the subsamples. Such fluctuations may occur rarely but, when they do occur, their magnitude can be extremely high (see Tables 2 and 3). This becomes increasingly important when working with highly dependent expression levels rather than with elements of the sequence $\{\delta_i\}$. The same is true for power studies with small effect sizes.

TABLE 2
*The mean and standard deviation of the number of type 1 errors and the false discovery rate (FDR). Resampling experiment with the effect size of two standard deviations of the corresponding log-expression levels. Sample size: $n_1 = n_2 = 10$*

|  | $\delta$-sequence | **Expression levels** |
|---|---|---|
| False positives (mean) | 3.84 | 6.20 |
| False positives (std) | 10.28 | 82.11 |
| False positives (range) | [0, 238] | [0, 1939] |
| FDR (mean) | 0.010 | 0.006 |
| FDR (std) | 0.022 | 0.053 |

The mean and standard deviation of the number of false discoveries were 3.84 (compare with a nominal level of 9) and 10.28, respectively. Recall that there were a total of 3542 hypotheses tested in this resampling experiment. The mean and standard deviation of the FDR were equal to 0.010 and 0.022, respectively. The results appear to be remarkably stable to the initial (i.e., obtained from the first split) ordering of genes based on sample standard deviations.

TABLE 3
*The mean and standard deviation of the number of type 1 errors and the false discovery rate (FDR). Resampling experiment with the effect size of one standard deviation of the corresponding log-expression levels. Sample size: $n_1 = n_2 = 10$*

|  | $\delta$-sequence | **Expression levels** |
|---|---|---|
| False positives (mean) | 4.06 | 7.99 |
| False positives (std) | 3.82 | 45.87 |
| False positives (range) | [0, 116] | [0, 1021] |
| FDR (mean) | 0.015 | 0.013 |
| FDR (std) | 0.026 | 0.086 |

For comparison, we conducted a similar study directly with the original log- expression levels. To this end, all genes with even numbers constituting the 350 pairs identified earlier by gene ordering were selected and the standard deviations of their log-expression levels were estimated from the first split in the previous resampling experiment. The effect size was set at two standard deviations as before. The resampling procedure was identical to the one applied to the sequence $\{\delta_i\}$ in the first experiment. In this case, the mean and standard deviation of the number of false discoveries were equal to 6.20 and 82.11, respectively, given the same nominal control level. The mean and standard deviation of the FDR were equal to 0.006 and 0.053, respectively. Note that the standard deviation is an order of magnitude larger than the corresponding mean value for both the number of false discoveries and the FDR. The range of the random number of false positives is [0, 1939], which is much wider than that ([0, 238]) for the $\delta$-sequence even if the latter is assessed in terms of single genes rather than gene pairs. It is clear that there is still some residual weak correlation between elements of the $\delta$-sequence in the data, but its strength is beyond compare to that between the original expression signals.

The same experiment was conducted with smaller effect sizes. More specifically, the effect size was equal to one standard deviation for both the elements of the $\delta$-sequence and log-expression of individual genes. For the $\delta$-sequence, the mean and standard deviation of the number of false discoveries were equal to 4.06 and 7.99, respectively. For the genes, however, their corresponding values were 3.82 and 45.87. The mean and standard deviation of the FDR were equal to 0.015 and 0.026 for the $\delta$-sequence, as compared to 0.013 and 0.086 for the individual genes. Again, the effect manifests itself in the standard deviations, although its magnitude is not as high as in the previous example with larger effect sizes.

The performance indicators estimated from both resampling experiments are summarized in Tables 2 and 3. These tables show how complex the outcomes of multiple testing can be in correlated data because of mutual dependence between true and false discoveries. A pictorial comparison of the two selection processes in terms of their outcomes in all 3000 successive subsamples is given in Supplementary Material 5.

REMARK 4. The proposed procedure attempts to detect changes in a specific characteristic (increment) of nonoverlapping gene pairs, given a specific ordering of the initial list of genes. The conventional procedure is designed to detect changes in expression levels of all individual genes irrespective of their order. Strictly speaking, the two procedures are not directly comparable, as they are designed to perform different tasks. From our resampling study, it follows that the proposed procedure performs its own task better. However, this procedure also provides superior characteristics (means and variances) of the power and type 1 error rates when used for testing changes in expression levels of individual genes. This fact was established by a similar experimentation with the empirical Bayes method and will be discussed at length in another paper.

The above example shows that the recourse to the sequence $\{\delta_i\}$ provides a highly accurate control of type 1 errors when testing differential expression of nonoverlapping gene pairs. The outlook for the nonparametric empirical Bayes method is even brighter within the proposed framework and our preliminary studies have already confirmed this conjecture. It seems to us that the empirical Bayes methodology can uncover the potential of this approach to the fullest extent to allow investigators to work with small samples of subjects. From this perspective, the economic impact of our suggestion can be quite high and we intend to explore this possibility more closely in future publications.

At first glance, the price to be paid for taking advantage of the sequence $\{\delta_i\}$ of weakly dependent random variables in microarray data is that some differentially expressed individual genes may remain undetected. Such genes may still be hidden in those gene pairs that do not show a significant change in the associated increment $\delta$ in two-sample comparisons. Although events of this sort are expected to be rare, this raises a natural concern. However, our studies have shown that the opposite situation is observed in real data, that is, the increments $\delta_i$ are subject to much more pronounced changes across different phenotypes than the expression levels. When comparing the data on two types of leukemia (HYPERDIP and TELL), we computed the Kolmogorov–Smirnov statistics associated with expression signals of all genes. We found 5.1% of these statistics to exceed the critical level at a significance level of 0.05. When the statistics were computed for elements of the $\delta$-sequence, more than 38% were found in the critical region at the same significance level. A similar behavior of the $\delta$-sequence versus gene expression was seen in the PCMIT data. This phenomenon is likely to be attributable to a special correlation structure of microarray data. Regardless of the cause, it is clearly beneficial from the practical standpoint.

Another advantage of the sequence $\{\delta_i\}$ is that all its members are essentially free from the technical noise. This is obvious when considering the array-specific noise model given by formula (4). Additionally, the ordering of genes by the variance of their expression level makes the noise levels imposed on both members of each gene pair quite close to each other, thereby substantially reducing the noise component in $\delta_i$ even under the more general array-gene-specific model (3). Therefore, the statistical inference based on the $\delta$-sequence is practically unaffected by measurement errors.

**7. Concluding remarks.** The existence of distinct correlation substructures, as well as sequences of near-independent random variables, in gene expression data is of biological interest in its own right, regardless of their practical implications in the arena of statistical methods for microarray data analysis. Such structures can be discerned in real data providing an appropriate choice of gene ordering is made. Ordering by values of the variance of gene expression levels has proven to be very fruitful in this regard. We believe this is no accident because the inter-subject variability of gene expression is likely to reflect the tightness of control of a particular gene function by genomic regulatory mechanisms. Therefore, this way of ordering may be biologically meaningful. The proposed ordering seems also natural in view of specific types of dependence uncovered in pairs and triples of genes. This does not mean, however, that the chosen ordering is unique; other orderings of genes may well prove useful in future studies. Deciphering sufficiently simple correlation substructures in microarray data seems to be a much more promising approach than trying to make sense out of the sea of all gene pairs, to say nothing of triples or larger subsets of genes.

The multivariate structure of gene expression data is extremely complex. To get a plausible image of the processes of gene expression, one should think of a gigantic network of interacting genes, in which the functioning of every gene can be affected by products of a huge number of other genes. We probably will never have the ability to fully reconstruct such a network, at least not in the foreseeable future. However, much can be learned about the structure of this network and its quantitative characteristics by studying stochastic dependencies in small gene sets such as pairs or triples of genes. Even this relatively moderate step toward gaining a better insight into the multivariate structure of microarray data may revolutionize not only the practice of microarray data analysis, but also the way of conceptual thinking of gene function.

In the present paper we discuss the $\delta$-sequence of independent increments as an interesting biological phenomenon. This phenomenon may have far-reaching implications for statistical methodology of microarray data analysis. From this perspective, an important observation is that marginal distributions of the increments $\delta_i$ show significant changes when comparing two different phenotypes, and hence, they hold much promise as sensitive indicators of differential expression of gene pairs.

All multiple testing procedures are very unstable in the presence of correlations between test statistics. This instability manifests itself in a high variability of the number of false and true discoveries from sample to sample, thereby causing instability of the reported lists of candidate genes [Qiu, Klebanov and Yakovlev (2005b), Qiu et al. (2006), Gordon et al. (2007)]. This well-documented effect is exacerbated in small samples. Resorting to the $\delta$-sequence can remedy this difficulty, which otherwise looks insurmountable. There is a plethora of technical issues that arise from the idea to exploit nice properties of the $\delta$-sequence for the purposes of statistical analysis. Such issues should be addressed in conjunction with specific multiple testing procedures and ample room still remains for further methodological research. A great deal of rigorous testing needs to be done before an ultimate inferential procedure can be put to practical use.

**Acknowledgments.** We are grateful to the anonymous reviewer and the associate editor for thorough reviews.

Department of Probability and Statistics  
Charles University  
Sokolovska 83  
Praha-8  
CZ-18675  
Czech Republic  
E-mail: levkleb@yahoo.com

Department of Biostatistics  
and Computational Biology  
University of Rochester  
601 Elmwood Avenue  
Box 630  
Rochester, New York 14642  
USA  
E-mail: andrei_yakovlev@urmc.rochester.edu